\documentclass[a4paper,12pt]{article}
\usepackage{epsfig}
\usepackage{times}
\usepackage{graphicx}
\usepackage{amsmath}
\usepackage{color}
\ifx\pdfoutput\undefined
\DeclareGraphicsExtensions{.epsi,.eps}
\else
\DeclareGraphicsExtensions{.pdf}
\fi
   
\newcommand{\mycite}[1]{\cite{#1}}

\newcommand{\beq}{\begin{equation}}
\newcommand{\eeq}{\end{equation}}
\newcommand{\beqa}{\begin{eqnarray}}
\newcommand{\eeqa}{\end{eqnarray}}

\def\opone{\leavevmode\hbox{\small1\kern-3.8pt\normalsize1}}

\begin{document}	
	\baselineskip24pt \noindent{\Large\bf Corner states of light in photonic waveguides}\\[3mm]
	\noindent {\small \bf Ashraf El Hassan, Flore K. Kunst, Alexander Moritz, Guillermo Andler, Emil J. Bergholtz$^*$ \& Mohamed Bourennane}\\[2mm]
	\noindent {\it Department of Physics, Stockholm University, S-10691, Stockholm, Sweden}\\[2mm]

	
	\noindent \textbf{The recently established paradigm of higher-order topological states of matter has shown that not only, as previously thought, edge and surface states but also states localised to corners can have robust and exotic properties. Here we report on the experimental realisation of novel corner states made out of classical light in three-dimensional photonic structures inscribed in glass samples using femtosecond (fs) laser technology. By creating and analysing waveguide arrays forming two-dimensional breathing kagome lattices in various sample geometries, we establish this as a platform for corner states exhibiting a remarkable degree of flexibility and control. In each sample geometry we measure eigenmodes that are localised at the corners in a finite frequency range in complete analogy with a theoretical model of the breathing kagome. Here, the measurements reveal that light can be ``fractionalised", corresponding to simultaneous localisation to each corner of a triangular sample, even in the presence of defects. The fabrication method applied in this work exposes the advantage of using fs-laser writing for producing compact three-dimensional devices thus paving the way for technological applications by simulating novel higher-order states of matter}.  

\newpage
\section*{Introduction}
Over the last few decades topologically protected boundary states have been the subject of intense theoretical and experimental investigation in physical contexts ranging from their original habitat in solid state materials to more recent incarnations in mechanical and photonic metamaterials. Saliently, such states are featured on the ends, edges or surfaces of one-, two- or three-dimensional topologically non-trivial (meta)materials, respectively \mycite{hasan2010colloquium, qi2011topological, hosur2013recent}, where their robustness is facilitated by topological protection---a global property of the system---and as such is difficult to destroy. Beyond the fundamental significance of these states, their robust properties are also of great promise in the context of technological applications including quantum computers, high-performance electronics and efficient information transfer devices. 

In parallel, optical waveguide arrays have been at the forefront of experimental science. This line of research has now begun to make contact with the field of topological states on matter, and has been established as an exciting alternative platform in which topological phenomena can be realised \mycite{floquettopopt,lujoannopoulossoljacic,mukherjee2015observation}. In glaring contrast to their complex incarnations in the solid state, this new platform offers superior control and opportunities for exquisite design of desirable array structures. More precisely, these topological lattice models are engineered by carving waveguides in glass by making use of a femtosecond (fs) laser \mycite{davis1996writing}. The fs laser alters the refractive index in the three-dimensional photonic device, which leads to the creation of a path along which photons can propagate. Each two-dimensional slice perpendicular to the direction of the waveguides then forms a ``stationary" lattice, whereas the third dimension, the direction in which the guide is carved, is viewed as the time axis. This allows for the direct experimental simulation of a variety of single-particle phenomena, including the probing of topological bandstructures where the targeted states depend on the properties of the injected light. In particular, chiral edge states akin to those of Chern insulators have been observed by making use of helical waveguides \mycite{floquettopopt,maczewsky2017observation}, and topologically protected photonic mid-gap defect modes have been probed in distorted honeycomb arrays \mycite{noh2018observation}, whereas localised bulk modes have been reported in Lieb lattice arrays \mycite{mukherjee2015observation,lieb2}. 
 
Very recently, a paradigm shift in the theory of topological phases took place when it was realised that the presence of crystalline symmetries can lead to robust lower-dimensional  topological boundary states localised to corners of two- or three-dimensional lattices or the hinges of three-dimensional lattices \mycite{benalcazar2017quantized,langbehn2017reflection,song2017z,schindler2018higher,parameswaran2017topological,ezawa2018higher,kunst2018lattice}. These so-called higher-order topological phases are explicitly protected by the crystal symmetry and as such their protection does not carry the same robustness as in the case of ``ordinary" topological states. Instead they depend on subtle crystalline symmetries and can be naturally viewed as a variant of crystalline topological insulators \mycite{Fu2010}. The first experimental realisations of such higher-order boundary modes have very recently been achieved: the quadrupole topological insulator \mycite{benalcazar2017quantized} has been realised in topolectric circuits \mycite{lee2017topolectrical}, mechanical metamaterials \mycite{serra2018observation}, and microwave circuits \mycite{peterson2018quantized} while corner modes predicted to appear in breathing kagome lattices \mycite{ezawa2018higher,kunst2018lattice,xu2017topological,ezawa2018higher2,Akari} have been observed in the setting of acoustic metamaterials \mycite{xue2018acoustic,ni2018acoustic}.

In this work, we demonstrate a radically different realisation of the breathing kagome corner modes using the aforementioned optical waveguides inscribed in glass samples using fs laser technology. 
The structures are arranged in two different geometries, i.e., a rhombus and equilateral triangle shown in Fig.~\ref{fig:Schematicstructures}. The fabrication method applied in this work exposes the  advantage of using the fs laser writing to produce compact three-dimensional devices. Our measurements demonstrate that the light can be confined to corners/edges and guided with exquisite control by varying the relative waveguide separation and by tuning the wavelength of the light. These results are remarkably robust and persist in presence of defects, thus paving the way for novel technological applications of light based on robust corner states.  
In the remainder of this work we first describe the breathing kagome lattice model \mycite{ezawa2018higher} and its corner states \mycite{ezawa2018higher,kunst2018lattice}, then detail how this is realised within our waveguide setup, and present and elaborate on the experimental results. Technical details on the experiments are provided in the methods section. 

\section*{Breathing kagome lattice of light}
\begin{figure}[ht]
	\centering
	\begin{tabular}{ccc}
		\includegraphics[width=0.8\linewidth]{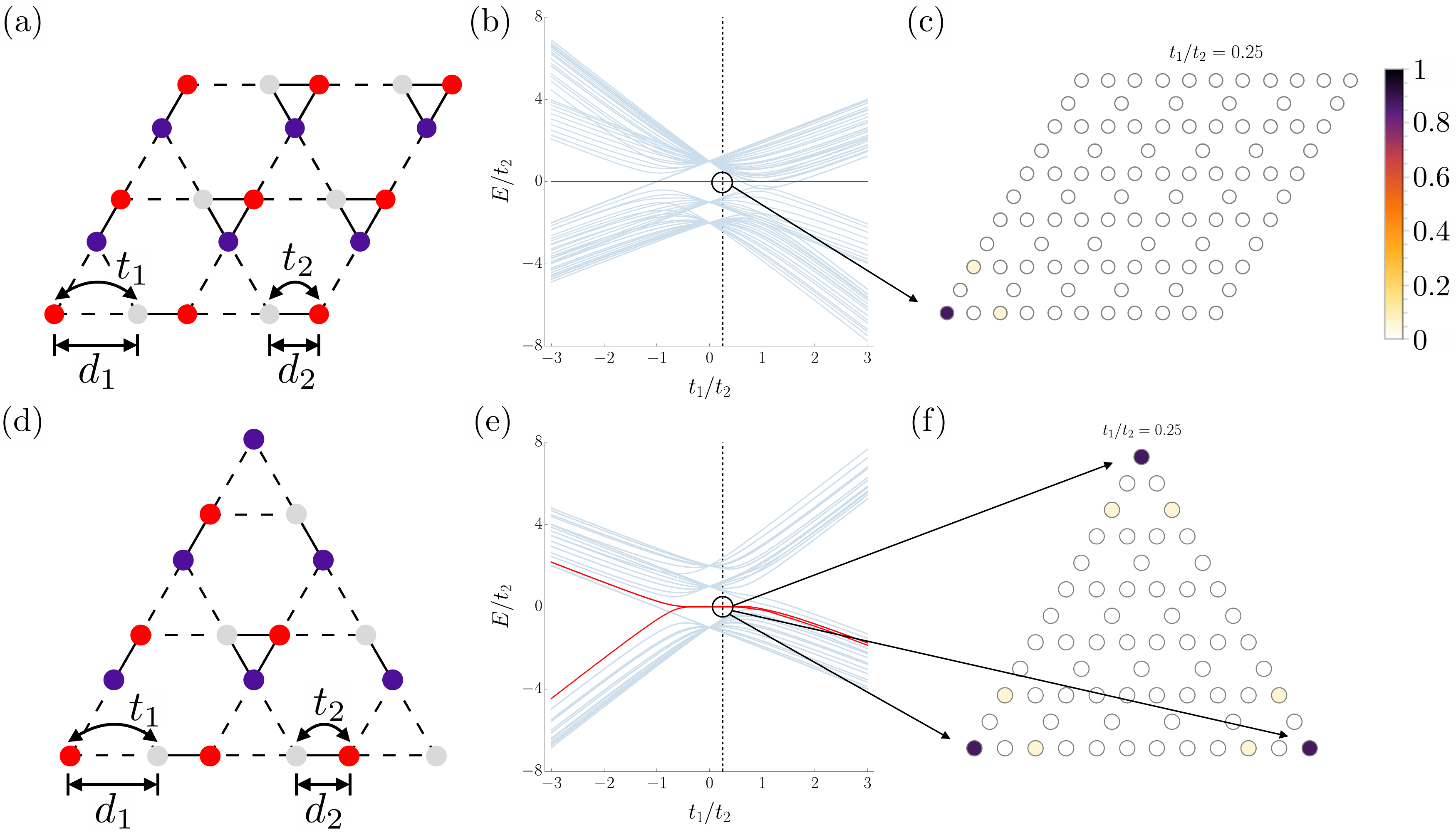}
	\end{tabular}
	\caption{Breathing kagome lattice. (a) and (d) show the lattice structure in the rhombic and triangular geometries respectively, with the two inequivalent tunnelling amplitudes, $t_1,t_2$ resulting from the two different lattice constants  $d_1,d_2$ as indicated. The corresponding theoretical energy spectra are shown in (b) and (e) while the total weight of the zero-energy wave functions are plotted in (c) and (f) for a representative parameter choice $t_1/t_2=0.25$ with the intensity bar in (c). 
	} 
	\label{fig:Schematicstructures}
\end{figure}

In the waveguide setup, Maxwells equations describing the propagation of light amounts to the paraxial equation \begin{equation}i\partial_z\Psi(x,y,z)=\left ( -\frac{1}{2k_0n_0}(\partial_x^2+\partial_y^2)-k_0\Delta n(x,y)\right )\Psi(x,y,z),
\end{equation} 
which is identical to the two-dimensional Schr\"odinger equation with the propagation direction $z$ playing the role of time $t$, and $\hbar=1$. The ``wave function" denotes the envelope of the electric field such that $E(x,y,z)=\Psi(x,y,z)e^{i(k_0z-\omega t)}$, and while the ``mass", $m=k_0n_0$, is set by the product of the wave number in free space, $k_0=n_0\omega / c$, and the refractive index, $n_0$, of the host material (the glass), the effective potential $V(x,y)=k_0\Delta n(x,y)$ is tailor-made by carving waveguides using the fs laser, which creates a strong spatial dependence of the local refractive index $\Delta n(x,y)$. In the limit of spatially sharp carving and weak evanescent coupling between the waveguides this system is accurately modelled by a tight-binding Hamiltonian whose hopping parameters depend on the setup and on the wavelength, $\lambda$, of the light. Here we use this technique to realise the breathing kagome tight-binding model which constitutes a network of corner sharing triangles as displayed in Fig.~\ref{fig:Schematicstructures}(a) and (d) with triangular and rhombic sample geometries, respectively. The lattice has two inequivalent lattice spacings, $d_1$ (dashed lines) and $d_2$ (solid lines) leading to two effective tunnelling strengths $t_i \sim e^{-d_i/\xi}$, where $\xi=\xi(\lambda)$ is set by the experiment and has the dimensions of length. We will for now proceed assuming that these two nearest-neighbour hopping parameters, namely $t_1$ and $t_2$, are sufficient to understand the breathing kagome system for a range of short enough wavelengths, which is fully consistent with the outcome of our experiments as discussed below. In this limit, the effective tight-binding Hamiltonian \mycite{ezawa2018higher} can thus  be written as 
\begin{equation}
H=t_1\sum_{\langle i,j\rangle\in \Delta}a^\dagger_ia_j +t_2\sum_{\langle i,j\rangle\in \nabla}a^\dagger_ia_j ,\label{tight}
\end{equation}
where the two sums are over the neighbours in up- and down-triangles as indicated by the solid and dashed lines in Fig.~\ref{fig:Schematicstructures}(a) and (d).
This setup naturally supports zero-energy states exponentially localised at corners \mycite{ezawa2018higher} as can be intuitively understood by thinking of the system in terms of a collection of judiciously combined Su-Schrieffer-Heeger (SSH) chains \mycite{su1979solitons}. The SSH chain is one of the first and most elementary examples of a symmetry-protected topological system and consists of a one-dimensional tight-binding hopping model with alternating hopping strengths $t_1$ and $t_2$, in close analogy with our kagome setup. For an even number of sites the SSH chain has robust zero energy states at each end of the chain when $|t_1/t_2| < 1$ assuming that $t_1$ is the hopping amplitude to the first and last site of the chain. For an odd chain there is always one zero-energy state, which is located on the end terminating with the weaker hopping amplitude. In direct analogy, the breathing kagome triangle has zero-energy states at each of its three corners in the parameter region $|t_1/t_2| < 1$ (cf. Figs.~\ref{fig:Schematicstructures}(e) and (f)) while the rhombus always support a corner state whose location is determined by which of the hopping amplitudes is larger (cf. Figs.~\ref{fig:Schematicstructures}(b) and (c)). 
An intuitive understanding for this behavior can be obtained from considering the dimerised $|t_1/t_2| << 1$ limit where the corner states simply correspond to unpaired sites. Away from this limit the corner state on the rhombus can be obtained exactly as  \begin{equation} |\Psi_{{\rm corner}}\rangle = \mathcal{N} \sum_{m,m'}\left(-t_1/t_2\right)^{m+m'} a^\dagger_{A,m,m'}|0\rangle , \label{cornerstate} \end{equation} where $a^\dagger_{A,m,m'}$ creates a state on the $A$ sublattice (red in Fig.~\ref{fig:Schematicstructures}(a)) in the unitcell numbered by $(m,m')$ starting with $(0,0)$ at the lower left corner \mycite{kunst2018lattice, kunst2018boundaries}. In addition to the corner localisation, a second striking feature of this state is that it resides on the (red) $A$ sublattice. While no longer exactly, analytically solvable in the finite triangular geometry, the corner states there are, to an exponentially good approximation in the large system limit, obtained in the $|t_1/t_2| < 1$ region by the three $C_3$ invariant combinations of the rhombus states. In contrast to the rhombus these now have finite weight on each of the three sublattices (red, grey, blue) since the corner sites live on different sublattices, and the corresponding energies exhibit a small finite size splitting. 

The dimerised limit also explains why there are zero-energy states located at the 60-degree corners while they are apparently absent for the wider 120-degree corners that occur in the rhombic geometry, as in the latter case the corner sites are connected by a strong bond. Away from this limit, this contrasting behaviour is related to the fact that, following the perimeter of the lattice there are domain walls, well known to trap soliton states in the corresponding SSH chains, at the narrow corners while the wide corners do not feature domain walls and thus no localised states. In the field theory language this corresponds to a sign change of the mass term in the narrow corners while the mass keeps the same sign in the case of the wider 120-degree corners. Consequently at the 120-degree corners there are no in-gap states.

\section*{Experiment}

\begin{figure}[ht]
	\centering
	\begin{tabular}{cc}
		\includegraphics[width=1.0\linewidth]{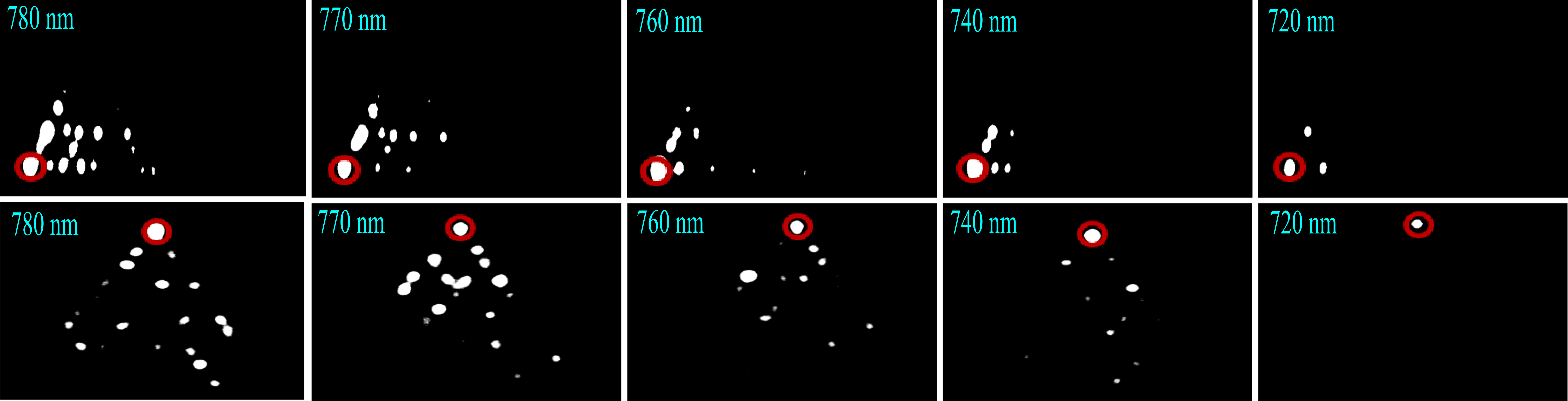}
	\end{tabular}
	\caption{Observation of light output from the rhombus (upper images) and triangle (lower images) lattices of waveguide arrays with $ d_1 = 12 \mu m$ and $d_2 = 7 \mu m$. CCD camera pictures of light emerging at the output facet of the rhombus and triangle waveguide arrays.
		The coherent input light at different wavelengths, $\lambda$, ranging from $780$ $nm$ to $720$ $nm$ is injected
		into the waveguide at the corner indicated by a red circle. The light propagation was $5$ $cm$.}
	\label{fig:Wavelengthscaning}
\end{figure}

We now proceed to discuss the implementation of the aforementioned lattice model using optical waveguide arrays. These arrays were written by a focused fs laser beam on $50$ $mm$ length glass samples that were mounted on 3 axis nano-positing stages. The waveguides were written such as to support a single mode at $780$ $nm$. The optical measurements were performed by exciting one waveguide on one facet of the glass sample by a tunable coherent light source. The observation of the emerging light on the opposite facet was done with the help of an optical objective and CCD camera (see the method section for the fabrication and optical measurements details.) 
To make contact with the theoretical breathing kagome model we begin by injecting coherent light in corner waveguides at different wavelengths and observe the emerging light pattern on a CCD camera. We set  $ d_1 = 12 \mu m$ and $d_2 = 7 \mu m$ such that in the limit of well localised light (small $\xi(\lambda)$) we have $t_1 < t_2$ and all other terms negligible. In this setup we hence expect a localised corner state in the lower left corner of the rhombus geometry, and in similarly each corner of the triangle in a finite wavelength window (at small $\lambda$). At wavelengths away from this window the effective tight-binding model is no longer accurately described by Eq. (\ref{tight}), hence there is no reason to expect localised corner states and we consequently expect to observe an essentially random light pattern. The measured light intensity distributions, fully confirming this picture, are displayed in Fig.~\ref{fig:Wavelengthscaning} for the rhombic (upper panels) and triangular (lower panels) lattices, where the corner states are very clearly resolved at  $\lambda=720$ $nm$ while being absent at longer wavelengths. 

\begin{figure}[ht]
	\centering
	\begin{tabular}{cc}
		\includegraphics[width=1.0\linewidth]{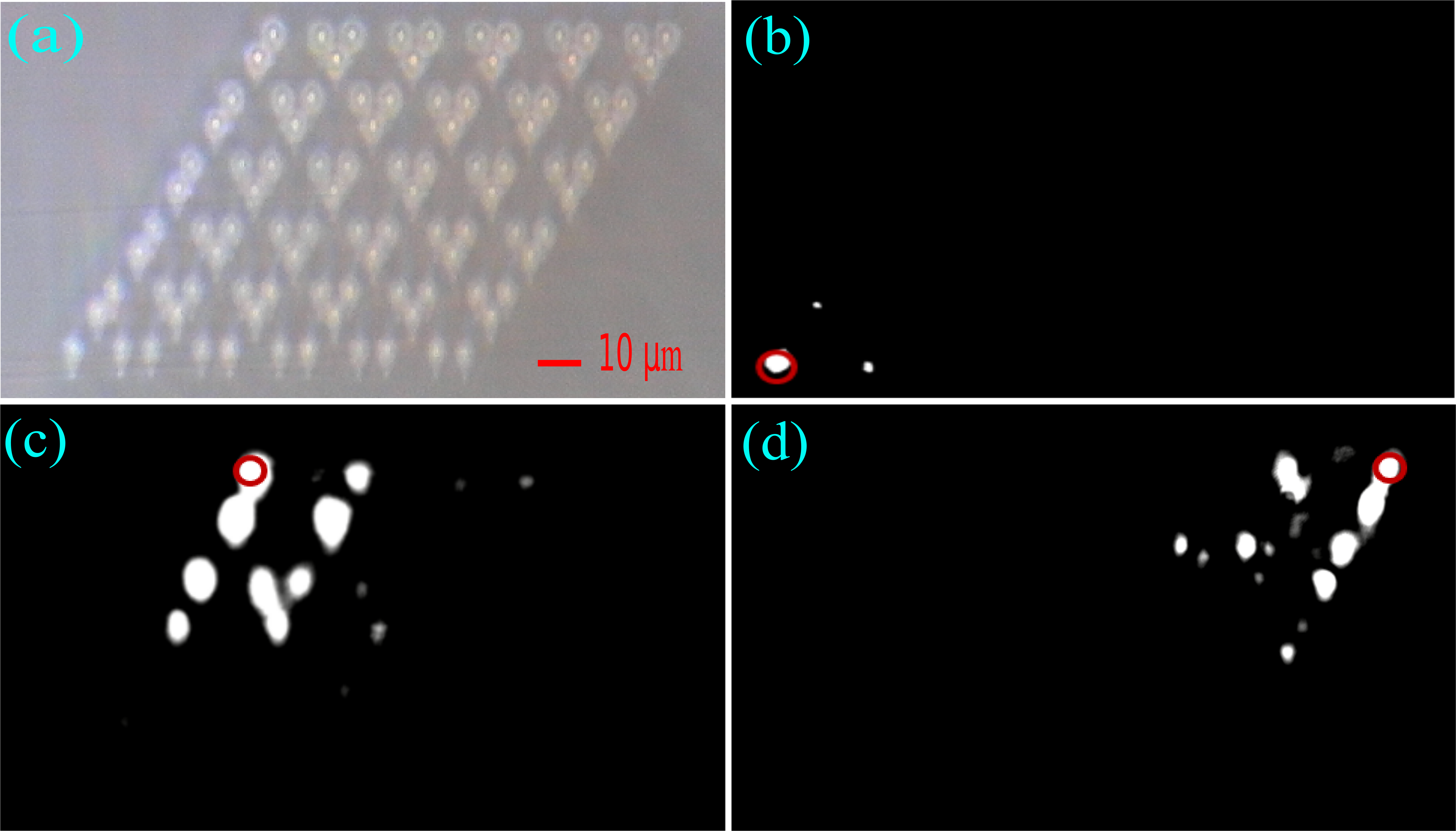}
	\end{tabular}
	\caption{Observation of the corner states in the rhombus lattice of waveguide array with $d_1 = 12 \mu m$ and $d_2 = 7 \mu m$. (a) Microscope image of the rhombus lattice. (b), (c), and (d) are the CCD camera pictures of light emerging at the output facet of the rhombus lattice of waveguide array. A coherent light at wavelength $720$ $nm$ is injected into the waveguide at the corner indicated by a red circle. The light propagation was $5$ $cm$. The microscopic picture (a) is in 1:1 scale with the experimental pictures (b), (c), and (d).}
	\label{fig:RompResults}
\end{figure}

In Fig.~\ref{fig:RompResults}, we fix the wavelength to $\lambda=720$ $nm$ and inject light at the three different types of corners of the rhombus. This reveals that the light stays localised if and only if the breathing kagome model hosts a corresponding corner state. When this is the case, the observed light (Fig.~\ref{fig:RompResults}(b)) is quantitatively very well described by the corner state wave function in Eq.~(\ref{cornerstate}) which has the tell-tale signature that, in addition to being exponentially localised to the corner, it  resides exclusively at a single sublattice while destructive interference makes it vanish on the other sublattices.

\begin{figure}[ht]
	\centering
	\begin{tabular}{cc}
	\includegraphics[width=1.0\linewidth]{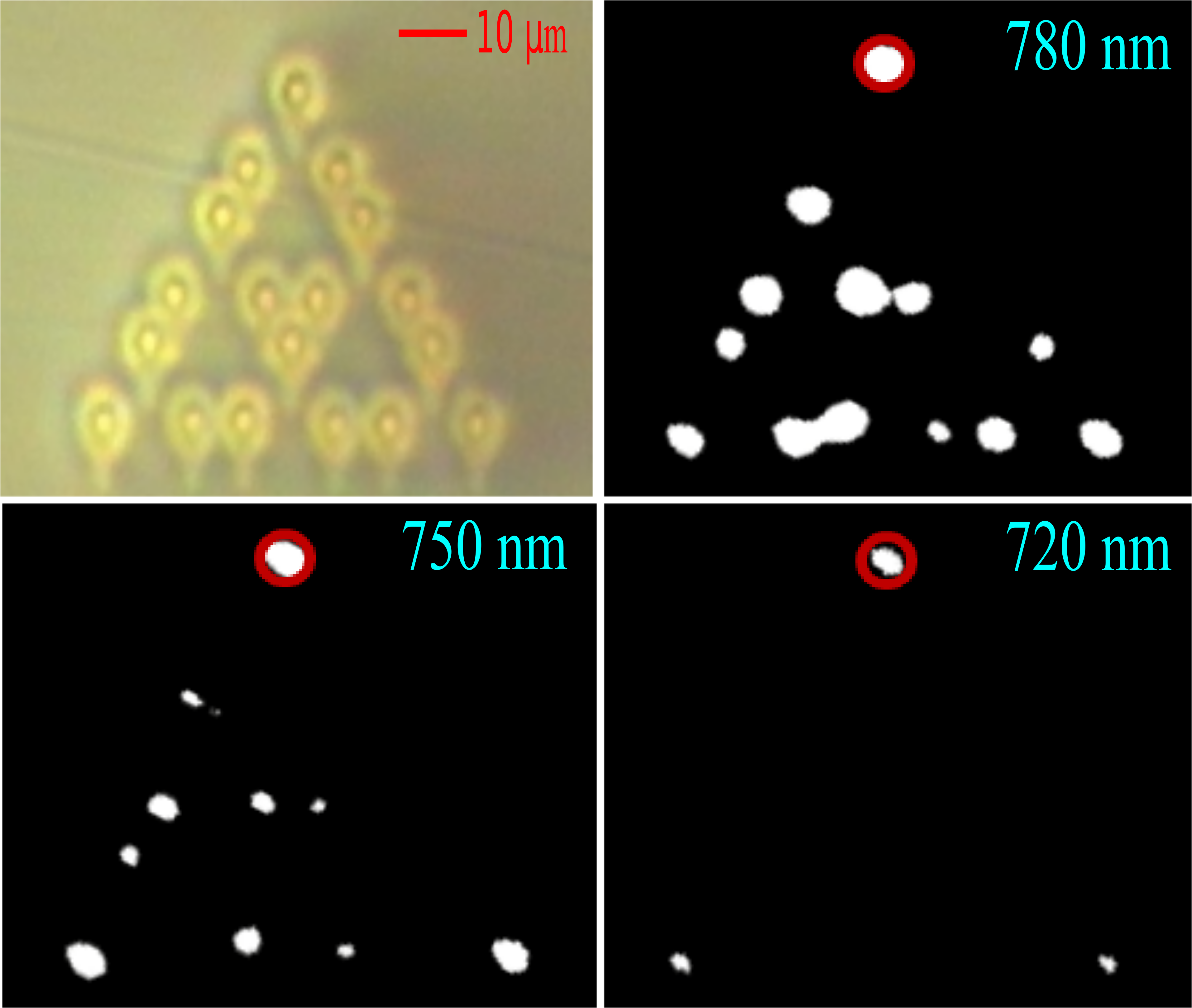}
	\end{tabular}
	\caption{Observation of the "fractionalised" corner states in the triangle lattice of waveguide array with $d_1 = 11 \mu m$ and $d_2 = 6 \mu m$. (a) Microscope image of the triangle lattice. (b), (c), and (d) are the CCD camera pictures of light emerging at the output facet of the triangle lattice of waveguide array. A coherent light at wavelengths varying from $780$ $nm$ to $720$ $nm$ is injected into the waveguide at the corner indicated by a red circle. The light propagation was $5$ $cm$. The microscopic picture (a) is in 1:1 scale with the experimental pictures (b), (c), and (d).}
	\label{fig:Triangleresults}
\end{figure}

\begin{figure}[ht]
	\centering
	\begin{tabular}{cc}
	\includegraphics[width=1.0\linewidth]{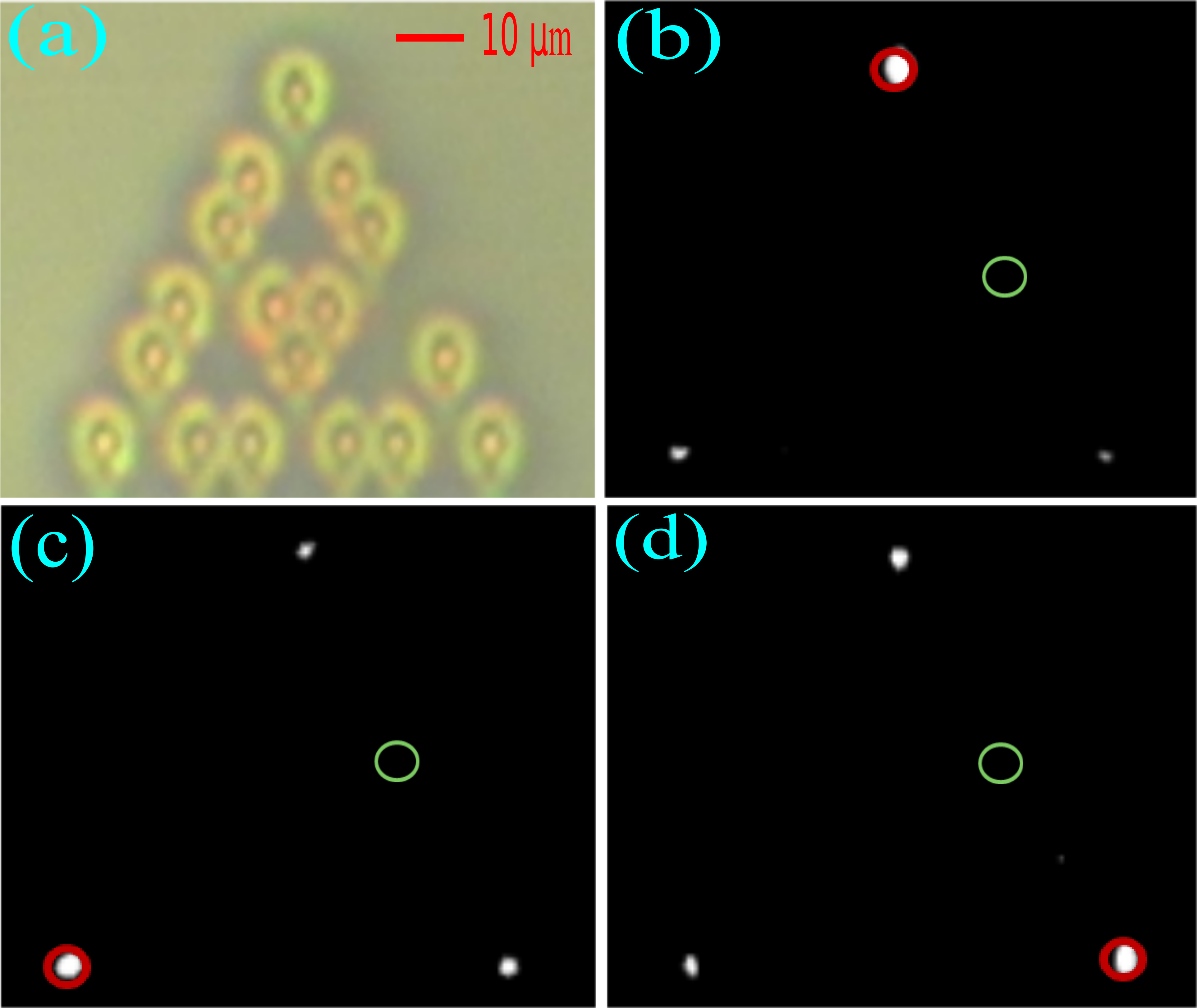}
	\end{tabular}
	\caption{Observation of the "fractionalised" corner states in a triangle lattice with a defect of a waveguide array with $d_1 = 11 \mu m$ and $d_2 = 6 \mu m$. (a) Microscope image of the triangle lattice. (b), (c), and (d) are the CCD camera pictures of light emerging at the output facet of the triangle lattice of waveguide array. A coherent light at wavelength $720$ $nm$ is injected into the waveguide at the corner indicated by a red circle and the defect, the missing waveguide, is marked by a green circle. The light propagation was $5$ $cm$. The microscopic picture (a) is in 1:1 scale with the experimental pictures (b), (c), and (d).}
	\label{fig:Triangledefectresults}
\end{figure}

An even more striking feature is revealed by studying the triangle geometry further. Injecting the light at a single corner formally corresponds to forming an equal amplitude superposition of three finite size eigenstates, each being close to the $C_3$ invariant combinations of the rhombus corner states, but whose energies are split at finite size. Thus, as we here probe the time evolution, the dynamically acquired phases imply relative phase shifts, $\Delta \phi=\Delta E\times t$, that should in theory generically lead to light at all corners given that either the time, i.e. the length of the waveguides, or the energy splitting, $\Delta E$, is large enough to observe appreciable relative phase shifts $\Delta \phi$. By studying waveguide arrays with smaller lattice spacing and fewer waveguides, which naturally have larger $\Delta E$, we are clearly able to observe such ``fractionalisation" of the light to all three corners of the triangle sample geometry (Fig.~\ref{fig:Triangleresults}) at $\lambda =720$ $nm$. As illustrated in Fig.~\ref{fig:Triangledefectresults}, this holds even in the presence of defects where this phenomenology is also shown to be robust with respect to the corner at which the light is injected. 

We have further corroborated our findings by studying several differently prepared samples, several of them including significant imperfections, yet they show that the appearance of corner states in a finite frequency window is a remarkably robust feature. This is related to the fact that, although there are local operators that immediately remove the corner states from zero energy, there are (near) zero-energy corner states similar to Eq. (\ref{cornerstate}) under more general conditions than one may naively expect. This in turn relates to the corner states' provenance from local destructive interference rather than from fine-tuning \mycite{kunst2018lattice}, and that the interference is a salient feature of the kagome lattice when distorted or even partially depleted.

\section*{Conclusion}
In this work we have provided the first realisation of kagome based corner states in an optical waveguide setup. In fact, the breathing kagome model that we have realised is remarkably simple, requiring only straight waveguides. This is in glaring contrast to the much more intricate setups with spiralling optical waveguides needed to realise ordinary first-order topological states. It also contrasts the recent beautiful albeit more complicated photonic realisation of protected mid-gap states located near the corners of a distorted honeycomb structure which require auxiliary waveguides to excite the zero-energy modes \mycite{noh2018observation}, whereas in our setup directly injecting the light at the corner waveguide suffices. Moreover, the ``fractionalisation" of light observed in the triangle geometries is a striking new feature of our work. Thus our setup does not only provide the first optical realisation of novel physical phenomena, its implementation is also remarkably simple and stable, hence drastically enhancing the prospects of incorporating these ideas in photonic technological devices.   

{\it Note added:} Independent of this work, other photonic realisations of corner states were simultaneously reported, albeit in two-dimensional setups \mycite{2dphoton,2dphoton2,2dphoton3,2dphoton4}, which are entirely different from ours using three-dimensional coupled waveguides.

\section*{Methods}
\subsection*{Fabrication of waveguide structures}
The topological photonic three-dimensional waveguide lattice structures were fabricated using a pulsed fs laser (BlueCut fs laser from Menlo Systems). The fs laser produces light pulses centered at a wavelength of $ 1030$ $ $$nm$, having a  duration of $ 350 $ $ fs$. In this experiment, we have used a repetition of $ 1 $  $MHz$. The waveguides were written in a Corning EAGLE2000 alumino-borosilicate glass sample with dimensions ($L= 50, W = 25, h =1.1$) $ mm $. To inscribe the waveguide structures, pulses of $ 210$ $ nJ$  were focused using a $ 50 $ $X$ objective with a numerical aperture (NA) of $ 0.55$. The waveguides were written at a depths between $70$ to $175$ $ \mu $$ m $ under the surface according to the designed structure, while the sample was translated at a constant speed of 30 mm/s by a high-precision three axes translation stage (A3200, Aerotech Inc.). The fabricated waveguides support a Gaussian single mode at $780$ $nm$, with a mode field diameter ($1/e^2$) of approximately $ 6-8 $ $\mu $$m$. 
The propagation losses were estimated to be around $ 0.3$ $dB/cm $ and the birefringence in the order of $ 7 \times 10^{-5}$.
After the structures were written in the glass sample, the lateral facets were carefully polished down to an optical quality with a roughness of $ 0.1$ $\mu$$m$. The sample was examined using a microscope, where each individual waveguide input, output, and position could be verified in accordance with the designed structure. 
\subsection*{Measurement}
For the observation of the topological isolation of the waveguide structures, the coherent light beam from a tunable laser (Cameleon Ultra II, Coherent) was launched into the glass sample  using a $100\times$ objective of $0.9$ NA, which was sufficient for individual excitation of each waveguide composing the  structure, while the output light of the glass sample was collected with help of a $100\times $, high NA objective. A CCD camera was used to picture the image profile of each individual waveguide forming the topological structure.



\textbf{Acknowledgements}
This work is supported  by the Swedish research council (VR) and the Knut and Alice Wallenberg Foundation. E. J. B. is a Wallenberg Academy Fellow.\\
 
\textbf{Author Contributions}

E.J.B. initiated the research. F.K.K. and E.J.B. derived the theoretical results. A.E.H., A.M., and G.A. designed, carried out the experiment, and performed the data analysis. M.B. supervised the experimental part. E.J.B. and F.K.K wrote the main text. M.B. and A.E.H. wrote the experimental part. All the authors discussed the results and contributed to the final version of the manuscript.\\
  
\textbf{Competing interests statement} 

The authors declare that they have no competing financial interests.\\

\textbf{Correspondence}

Correspondence and requests for materials
should be addressed to E.J.B.~(email: emil.bergholtz@fysik.su.se).

\newpage

\newpage
\newpage
\newpage
\newpage

\end{document}